# Watt-level, all-fiber, ultrafast Er/Yb-codoped double-clad fiber laser mode-locked by reduced graphene oxide interacting with a weak evanescent field

Lei Gao, Tao Zhu,[*] and Yujia Li

*Key Laboratory of Optoelectronic Technology & Systems (Ministry of Education), Chongqing University, Chongqing 400044, China*
[*]*Corresponding author: zhutao@cqu.edu.cn*

We propose a Watt-level, all-fiber, ultrafast Er/Yb-codoped double-clad fiber laser passively mode-locked by reduced graphene oxide (rGO) interacting with a weak evanescent field of photonic crystal fiber (PCF). The rGO solution is filled into the cladding holes of the PCF based on total reflection, and after evaporation, the rGO flakes bear only $1/10^7$ of the total energy in laser system, which enhances the thermal damage threshold and decreases the accumulated nonlinearity. By incorporating the saturable absorber into an Er/Yb-codoped fiber ring cavity, stable conventional soliton with a duration of 573 fs is generated, and a average output power up to 1.14 W is obtained. © 2015 Optical Society of America

High power, all-fiber, passively mode-locked fiber laser (ML) has draw tremendous interests for the advantages of free of alignment, excellent heat radiation, compactness, and output-coupling convenience, which has wide applications in nonlinear optics, micro-fabrication, medicine imaging, and precision metrology [1-4]. Conventional high power ML based on large mode area fiber [5] requires fundamental mode engineering such as high order filtering, and the bulky construction and complex coupling alignment in schemes based on free space optics is inconvenient [6]. Limited by damage threshold and accumulated nonlinearity, the average power in all-fiber ML systems based on semiconductor saturable absorber mirror [7], carbon nanotubes [8], topological insulator [9,10], and graphene [11-15] is only several or tens of mWs, so post-amplification must be employed to get high power ultrafast laser [16]. For the frequently used fiber ferule method, where light passing through graphene deposited on the surface of the fiber end face [12], the thermal accumulated burns the materials when the power in the resonator is high enough.

To increase the damage threshold, schemes based on taper/side-polished fiber that interacts with graphene have been proposed [16-19], where double-clad rare-doped fiber with large gain is utilized, and Q-switched lasers centered at 1 μm [20], 1.5 μm [21], and 2 μm with average power of tens of mWs were produced [4,22]. However, the period and pulse duration of Q-switched laser are sensitive to pump power variation, and the peak power is severely limited by the Q-switched operation, which even disappears when the saturable absorber (SA) is bleached at high pump power [21]. Besides, harmonic mode-locking (HML) can easily arise in the ML systems based on the evanescent-interacting methods, due to the enhanced accumulated nonlinearity when light is circulating thousands of times in the fiber cavity [17, 23], where graphene is interacting intensely with intense evanescent field provided by the taper/side-polished fiber. For example, the light intensity ratio of the fiber center and the place where graphene is deposited, κ, is about ~30 in [23], indicating that a large portion of light interacts with the materials each time when passing by the fiber.

It is feasible to decrease nonlinearity via attenuating the evanescent filed intensity by increasing the fiber taper diameter or reducing the polishing part of side-polished fiber. Although the interacting intensity ratio is reduced, the circulating numbers of light is increased to reach a steady state when pumped by a larger power [24]. Meanwhile, their thermal damage thresholds are much larger than that of the fiber ferule method. Our experimental results prove that stable mode-locking could be obtained even the intensity ratio, κ, is increased to ~ $10^7$. Yet, it is difficult to produce a sufficient long, uniform fiber taper with desired diameter, and the polarization-dependent loss of side-polished fiber causes significant nonlinear polarization rotation effect for mode-locking [25]. Moreover, as the evanescent fields of the two kinds of fiber are directly open to the environment, special package must be employed, and the present reports about high power ultrafast lasers based on taper/side-polished fiber are mainly Q-switched fiber laser.

Photonic crystal fiber (PCF) that supports endlessly single-mode operation provides a better way to solve the problems. The air holes in PCF facilitate the filling of nanomaterials, and once the sample is spliced in PCF, the closed space makes the mode-locking process robust to the environment disturbance, such as air flow. There are several reports about MLs based on graphene and graphene oxide solution filled in PCF [26-28]. Nevertheless, direct filling of claddings based on photonic-bandgap effect distorts the guiding mode of PCF, and the graphene oxide solution filled in PCF makes the mode-locking unstable. Furthermore, the nonlinearity is also accumulated as the intensity ratio intensity ratio, κ, is relative high in those schemes, which restricts the high power operation of stable ML with fundamental frequency.

In this letter, we demonstrate a Watt level, ultrafast ML based on PCF interacting weakly with reduced graphene oxide (rGO). The SA is produced by splicing PCF based on total reflection that filled by rGO flakes with single mode fiber (SMF, Corning SMF-28). The rGO flakes instead of rGO solution sealed in PCF make the mode-locking rather stable. Most importantly, as the light intensity of the place deposited with rGO is much smaller than that of the fiber center, the rGO flakes experience only $1/\kappa$ of the total intensity of the laser. That's to say, the laser energy is dispersed so that the rGO, which contributes saturable absorption property and possess ultrahigh nonlinearity coefficient in this experiment, bears a much smaller power density. Thus, the thermal damage threshold is increased and also the nonlinearity can be decreased to suppress the HML. We incorporate the SA into an Er/Yb-codoped double-clad fiber ring cavity, and a stable femtosecond duration conventional soliton with a maximum average power of 1.14 W is obtained.

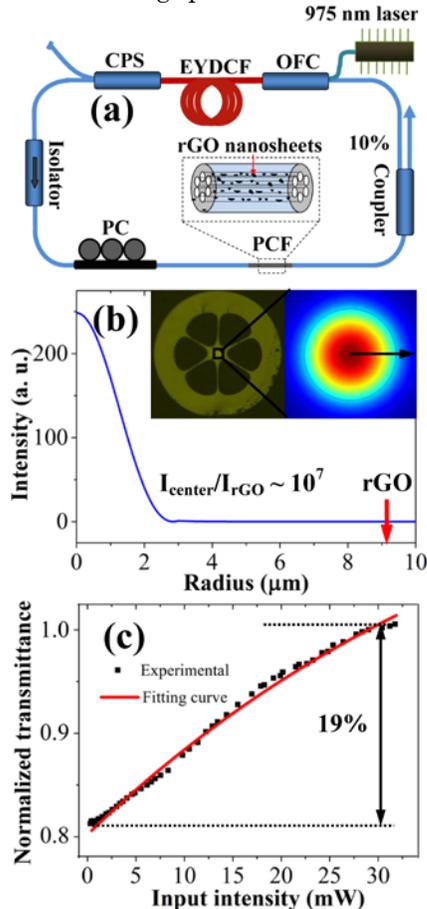

Fig.1 (a), Schematic setup of the fiber laser. (b), Light intensity vs fiber radius, and the cross section of PCF and its stimulated electric distributions of $LP_{01}$ mode are represented in the insets. (c), Normalized nonlinear transmission of the SA.

The setup of fiber cavity is shown in Fig. 1 (a), in which 8 m Er/Yb-codoped double-clad fiber (EYDCF, SM-EYDF-6/125-HE) is forward pumped by multimode 975 nm laser with a maximum power of 7 W through a optical fiber combiner (OFC). Due to the nonradiative cross relaxation between Yb and Er ions and the prevention of Er clusters, the EYDCF possess a higher pumping efficiency than that of conventional fiber, and the double-clad structure makes the fiber handle with a much higher pump power as a major portion of the pump light could be absorbed in the core of the active fiber. To prevent the thermal accumulation of the excess pump laser that can not be absorbed, a cladding power stripper (CPS) is used to strip the light propagating in the outer cladding of the gain fiber. The unidirectional operation is forced by a high power polarization independent isolator. The polarization state is optimized by a mechanical polarization controller (PC), and the output power is extracted by an optical coupler with 10% output. Besides, 10 m additional SMF is incorporated into the cavity, so the net anomalous dispersion cavity length of 18 m corresponds to a fundamental frequency of 11.73 MHz.

The laser output is visualized by an oscilloscope (Infiniium MSO 9404A, Agilent Tech.) together with a 350 MHz detector (PDB430C, Thorlabs Co,. Ltd), and the pulse duration is detected by an autocorrelator (APE, PulseCheck). The frequency domain information is analyzed by a frequency analyzer (Agilent, PSA E4447A), and an optical spectrum analyzer (Q8384, Advantest Corp.). To avoid the damage of those devices, a variable optical attenuator is inserted to change the output power.

The SA is produced by splicing 1.3 cm PCF whose cladding holes are filled by rGO dissolved in N,N-dimethylformamide with the Siphon Effect. The detail parameters of rGO are represented in [13,23]. Fig. 1 (b) depicts the cross section of PCF and its stimulated electric norm distributions of $LP_{01}$ mode, where light intensity vs fiber radius reveals that the intensity ratio, $\kappa$, is about $10^7$. Namely, only a very small percentage of light passing by the PCF will interact with rGO. This weakly interacting intensity enhances the thermal damage threshold and decreases the nonlinearity. To improve the stability of the mode-locking, we dry the rGO solution in PCF in 38° for 24 hours to avoid the nanosheets movement in solution. After evaporation, the PCF is spliced between SMFs with splice loss of ~3 dB.

We measure the nonlinear optical response of the SA based on a balanced two-detector methods as in [29]. The home-made ultrafast fiber laser centered at 1563 nm posses a duration of 285 fs and a repetition rate of 7.4 MHz, and a variable optical attenuator is used to change the average power. As the rGO flakes are only partially interacting with PCF, we use the pump power rather than the pulse intensity. The experimental data shown in Fig. 1 (c) is fitted by the typical nonlinear transmission formula, where an absolute modulation depth of 19% is obtained. This modulation depth could be increased by increasing the length of PCF.

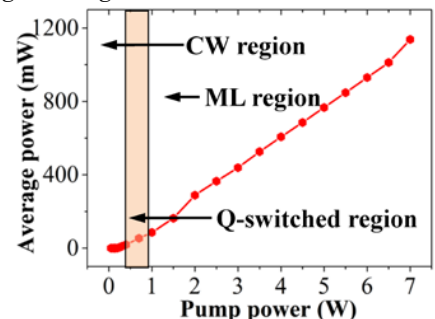

Fig. 2. The average output power under different pump powers, where CW region, Q-switched region, and ML region are shown, respectively.

As shown in Fig. 2, the threshold of the continuous wave (CW) laser operation is 0.175 W, and after PC optimization, Q-switched laser is formed when the pump power is increased to 0.35 W, and stable conventional soliton operation is achieved when the pump laser is 0.9 W. The typical optical spectrum of the Q-switched laser for pump power at 0.8 W centered at 1564.9 nm possesses a full width at half maximum (FWHM) of ~0.2 nm, and the pulse duration is about 4.36 μs. We find that the pulse period varies from 26.4 kHz to 37.1 kHz when the pump power changes from 0.35 W to 0.9 W.

Fig. 3 shows the typical characteristics of the ML region for pump power at 1.7 W. The optical spectrum in Fig. 3 (a) centered at 1564.1 nm posses a FWHM of 7.4 nm, and clear Kelly sidebands on both sides indicate it is a conventional soliton laser. The pulse duration deduced from its autocorrelation trace in Fig. 3 (b) is about 573 fs, corresponding to a time-bandwidth product (TBP) of 0.354. This value is slightly larger than that of a standard transform limited pulse, indicating a small chirp in the ML. The period of pulse train is 85.33 ns, and the RF spectrum shown in Fig. 3 (d) contains a fundamental frequency of 11.73 MHz with a contras ratio larger than 60 dB. We note that the intensities of the sidebands are unsymmetrical, of which may because the nonflat gain of the EYDCF, and the high peak power may also responsible for this asymmetry.

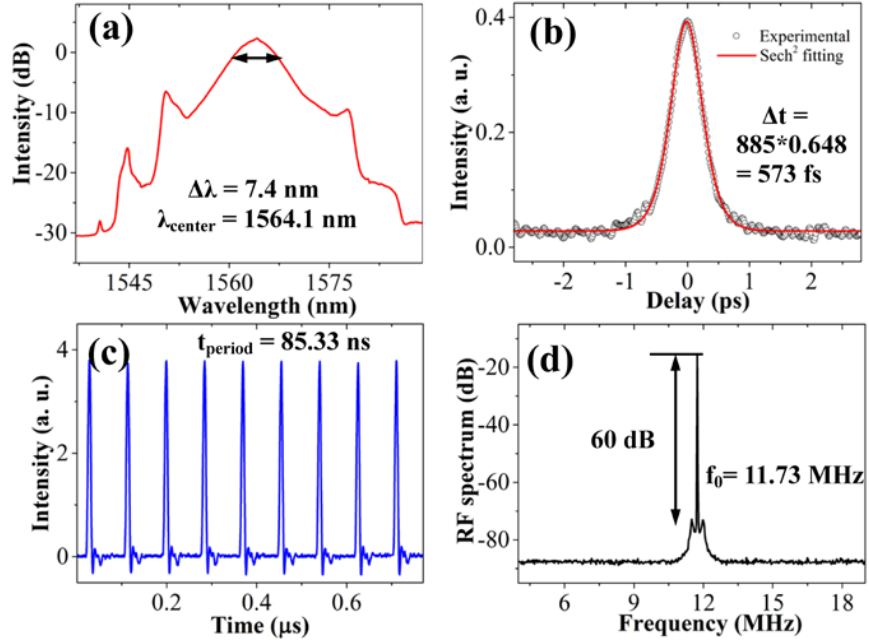

Fig. 3. Characteristics of conventional soliton for pump power at 1.7 W. (a) Optical spectrum, and (b) the corresponding autocorrelation trace. (c) Temporal pulse train, and (d) RF spectrum at the fundamental frequency of 11.73 MHz.

When the pump power is larger than 3 W, the mode-locking process deteriorates. Fig. 4 depicts the optical spectra for the pump power at 3.7 W and 7 W, respectively. When increasing the pump power, we keep the polarization state unchanged as in Fig. 3. As shown, CW portion appears on the top of the optical spectrum, and a narrower and bule-shifted optical spectrum indicates ML with a broader pulse duration. This deterioration of mode-locking is mainly induced by the large nonlinear phase change as the peak power of the ML is far larger than those of previous reports [12, 15]. The giant nonlinear phase variation imposed on the each longitudinal modes makes the mode-locking process unstable. Besides, the polarization-dependence of graphene is also responsible to the deterioration of mode-locking.

Although HML may occur for pump power larger than 5 W, we successfully obtain the conventional soliton with fundamental frequency by optimizing the polarization state via rotating PC mechanically, through which the gain/loss is adjusted. This fundamental frequency ML operation could be maintained for several hours in the lab condition. The optical spectrum for pump power at 7 W is represented in Fig. 4, which possesses a FWHM of 3.7 nm, and its pulse duration is about 1.32 ps. The TBP in this region is about 0.408, which is slightly larger than that for pump power at 1.7 W due to the more intense self-phase modulation.

In our experiment, the heat dissipation problem has to be taken into account for high pump power, as the laser system takes several seconds to achieve stable state. We find that this time increases with the increment of pump power, indicating that more circulating numbers in the resonator is required to get stable ML based on rGO interacting with a weak evanescent filed when pumped with a larger energy. In addition, no SA damage ever been find throughout the whole experiment, which is mainly because that only $1/10^7$ of the total energy in the laser system is interacting with rGO, making it an effective choice for high power ML.

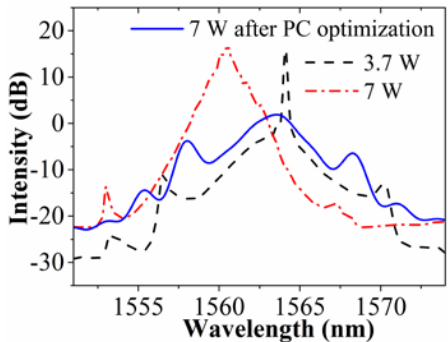

Fig. 4. Optical spectra for pump powers at 3.7 W and 7 W, respectively. The solid line represents the optical spectrum after PC optimization for pump power at 7 W.

In summary, we have proposed a Watt-level, all-fiber, ultrafast fiber laser based on EYDCF incorporating rGO interacting weakly with PCF. As the rGO nanoflakes are interacting with PCF based on total reflection by a weak evanescent intensity, the thermal damage threshold is highly enhanced, and the accumulated nonlinearity can be reduced to suppress HML. The usage of rGO rather than graphene solution makes the mode-locking rather stable. We obtained a stable conventional soliton with a duration of 573 fs, and achieved a maximum average output of 1.14 W for pump power at 7 W. This proves that the saturable absorption material interacting with a weak evanescent field is an attractive choice for high power ML. This kind of high power, all-fiber, ultrafast fiber laser may find potential applications in laser fabrication, nonlinear optics, medicine, and optical metrology.

This work was supported by Natural Science Foundation of China (No. 61405020, 61475029, and 61377066) and the Science Fund for Distinguished Young Scholars of Chongqing (No. CSTC2014JCYJJQ40002).